# Configurational Effects on Alfvénic modes and Confinement in the H-1NF Heliac

**B. D. Blackwell** [1)], D.G. Pretty [4)], J. Howard [1)], R. Nazikian [9)], S.T.A. Kumar[5)], D. Oliver [6)], D. Byrne [1)], J.H. Harris [7)], C.A. Nuhrenberg [8)], M. McGann[2)], R.L. Dewar[2)], F. Detering[1,2)], M. Hegland [3)], G.I. Potter [1)], J.W. Read [1)]

1) Plasma Research Laboratory, and
2) Department of Theoretical Physics, Research School of Physical Sciences and Engineering, and
3) Mathematical Sciences Institute, all of the Australian National University, ACT 0200,AUSTRALIA.
4) Present Address: Laboratorio Nacional de Fusión, EURATOM-CIEMAT, 28040 Madrid, Spain.
5) Present Address: Department of Physics, University of Wisconsin-Madison, USA.
6) Present Address: Research Group, Boronia Capital, Sydney Australia
7) Oak Ridge National Laboratory, Tn, USA.
8) Max-Planck-Institut für Plasmaphysik, Greifswald
9) Princeton Plasma Physics Laboratory, NJ, USA.

e-mail contact of main author:: boyd.blackwell@anu.edu.au

**Abstract:** The "flexible Heliac" coil set of helical axis stellarator H-1 (major radius R=1m, and average minor radius <r> ~ 0.15-0.2 m) permits access to a wide range of magnetic configurations. Surprisingly, in the absence of any obvious population of energetic particles, Alfvén modes normally associated with energetic populations in larger scale fusion experiments are observed. Using H-1's unique combination of flexibility and variety of advanced diagnostics RF-generated plasma in H-1 is shown to have a very complex dependence on configuration of both the electron density and the nature of fluctuations in the MHD Alfvén range. The magnetic fluctuations range from highly coherent, often multi-frequency, both simultaneously or sequentially, to approaching broad-band (df/f ~ 0.02-0.5), in the range 1-200 kHz. Application of datamining techniques to a wide range of configurations classifies these fluctuations and extracts poloidal and toroidal mode numbers, revealing that a significant class of fluctuations exhibit scaling which is i) Alfvénic with electron density (within a constant factor) and ii) shear Alfvénic in rotational transform. An array of optical and interferometric diagnostics is combined with the magnetic probe arrays to provide initial information on the internal structure of the MHD modes, and associated 3D effects. The configurational dependence is closely related to the presence of low order rational surfaces; density falls to very low values near, but not precisely at these rational values. Results from a uniquely accurate magnetic field mapping system, combined with a comprehensive model of the vacuum magnetic field in H-1 show that magnetic islands should not dominate the confinement of the configuration, and indicate that the strong dependence of plasma density on configuration may be a attributable to variations in plasma generation favouring the presence of islands.

## 1. Introduction

H-1 [1] is a medium sized helical axis stellarator of major radius R=1m, and average minor radius <r> ~ 0.15-0.2 m. Its flexible heliac[2] coil set (Figure 1) permits access to a wide range of magnetic configurations, both favourable and unfavourable, achieved by precise control of the ratio $k_h$ of the helical winding current to the ring coil current, and two sets of vertical field coils. This provides rotational transform $\iota$ in the range $0.9 < \iota_0 < 1.5$ for $B_0 < 1T$, shear in the transform of both the positive sign typical of stellarators and negative or tokamak-like, and magnetic well from ~5% to -2% (i.e. hill). RF-generated plasma shows a very complex dependence on configuration[3];

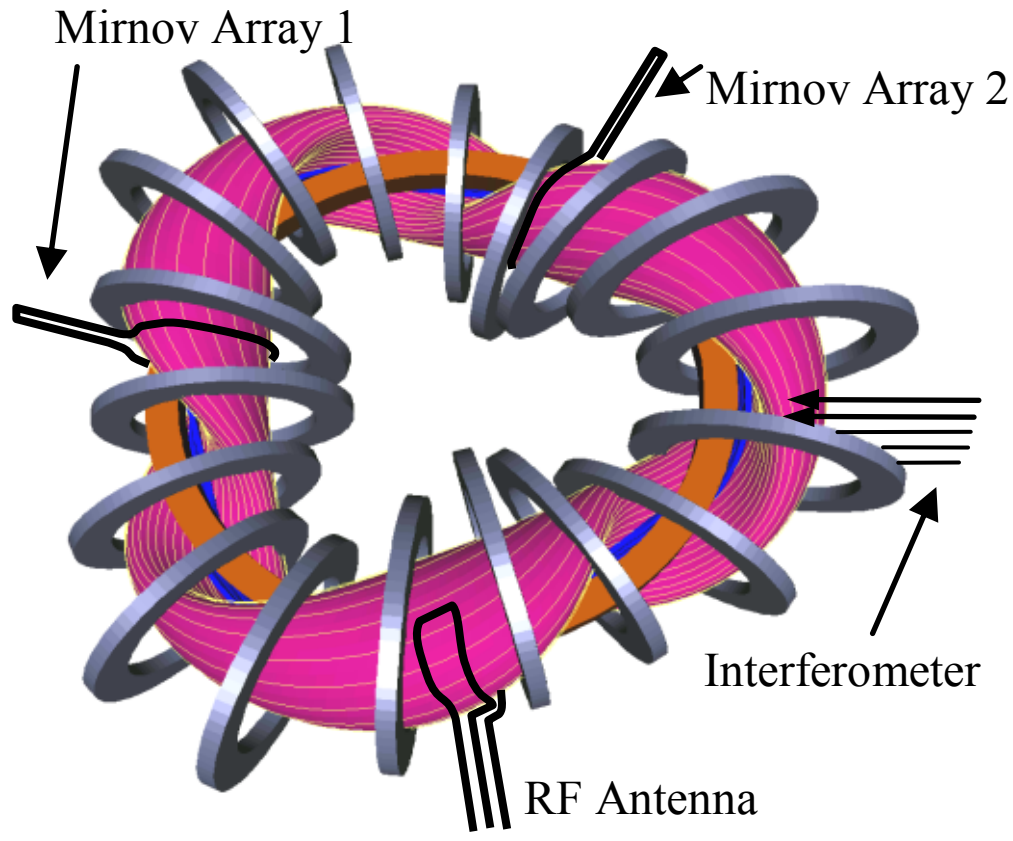

***Figure 1:*** *H-1 plasma showing location of Mirnov arrays, RF antenna and interferometer; and 18 of 36 TF coils*

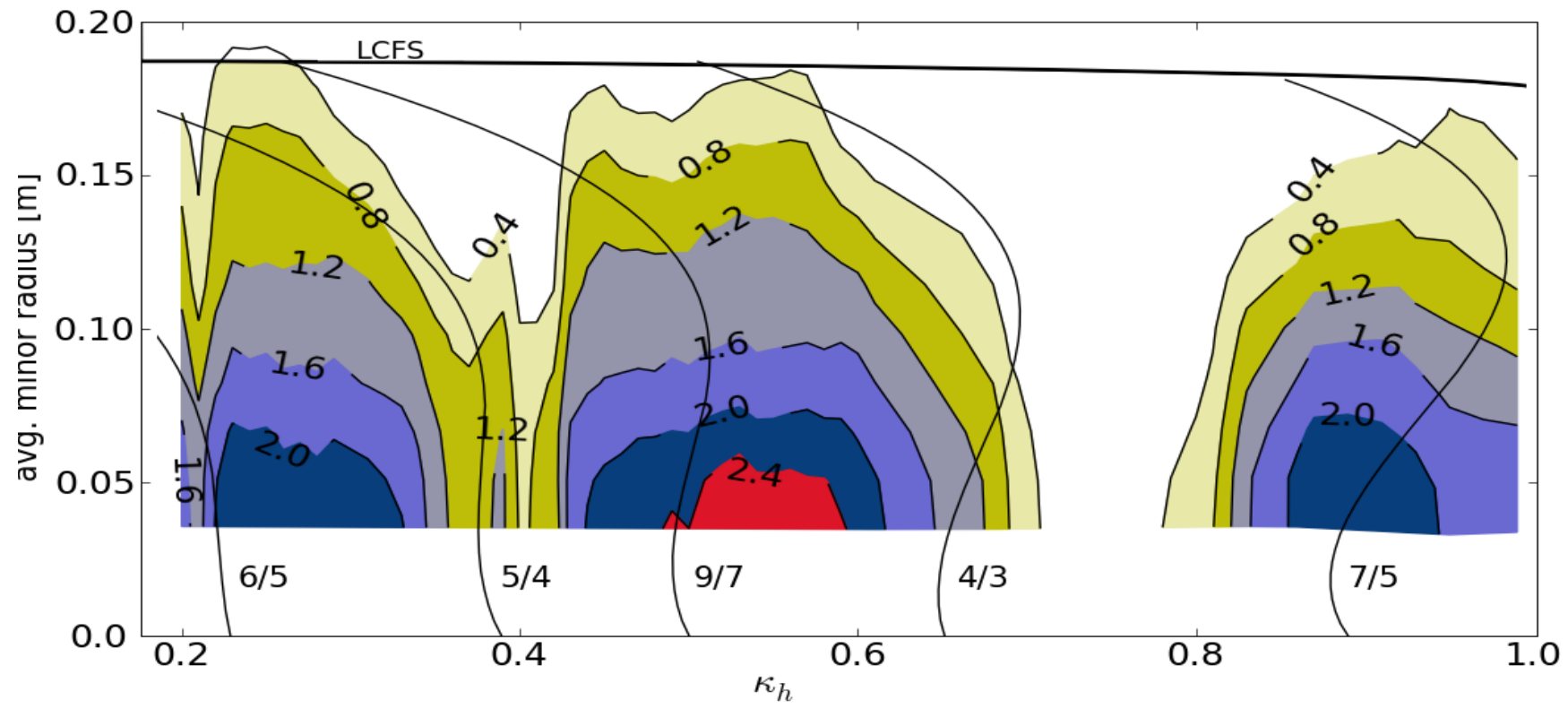

***Fig. 2:* a)** *Contours of plasma density radial profile as configuration is varied*

both the electron density (Figure 3) and the nature of fluctuations vary in a manner correlated with the presence of low order rational values of rotational transform. Under these conditions ( H/D/He mixtures, $B_0$~0.5, $n_e$~$10^{18}m^{-3}$) signals range from highly coherent, often multi-frequency in sequence or simultaneously, to approaching broad band (δf/f ~ 0.02-0.5), in the range 1-200 kHz, the higher frequency fluctuations (f>15 kHz) predominantly magnetic with amplitudes ~ 1gauss.

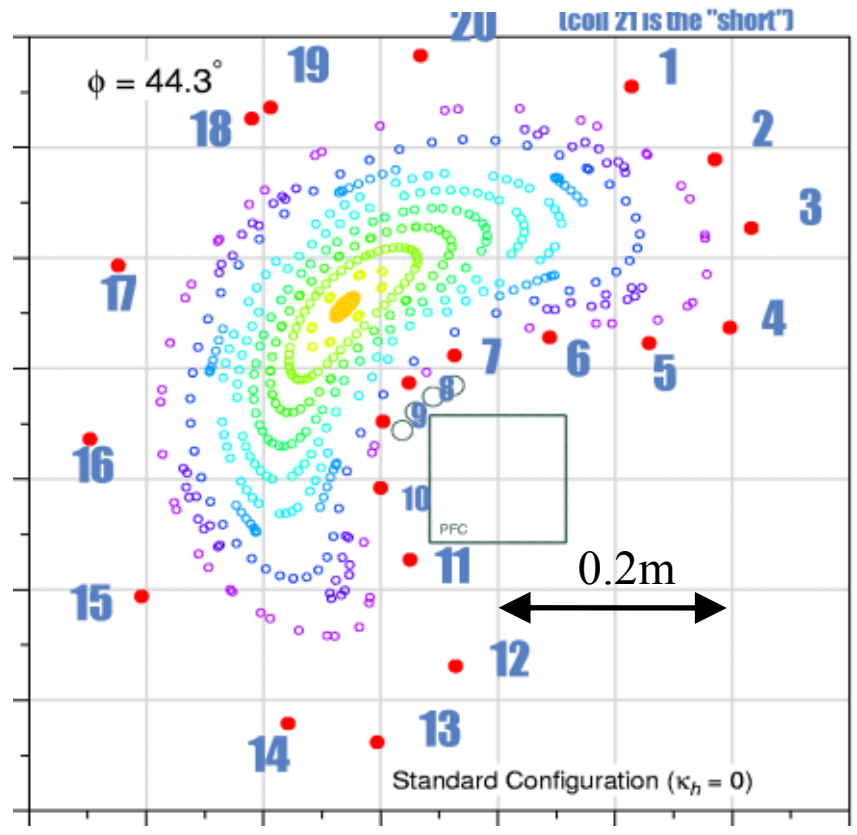

***Figure 3*** *Coil position (1-20) and plasma at* $\phi$ *=44.3°*

## 2. Magnetic Fluctuations

Data from two arrays (Fig 2a,b) of 20 magnetic probes and several other individual probes, for a series of ~80 magnetic configurations in the range 1.1 < $\iota_0$ < 1.4 provides much information about the nature of these instabilities, but amounts to a formidable data set. Data mining techniques[4] allow automated processing using Fourier and SVD[5] techniques, in the time domain and in space respectively, reducing the multi-channel timeseries data to a much smaller set of "fluctuation structures" on a much coarser time grid, characterised by a dominant frequency, amplitude, and relative phase of magnetic probe channels. At this level, data from hundreds of shots can be rapidly searched in an SQL database on a desktop computer, and datasets of the entire history of a device can be searched on a supercomputer.

Further data filtering typically involves entropy and/or energy thresholds, and classification of similar phenomena by clustering[6] in the multi-dimensional (15-40) space of phase difference between adjacent coils. This effectively groups according to mode numbers, both toroidal and poloidal without the need for spatial Fourier analysis. This is a significant advantage in the strongly toroidal geometry of compact devices, or in the three dimensional geometry typical of stellarators where mode structure is far from sinusoidal.

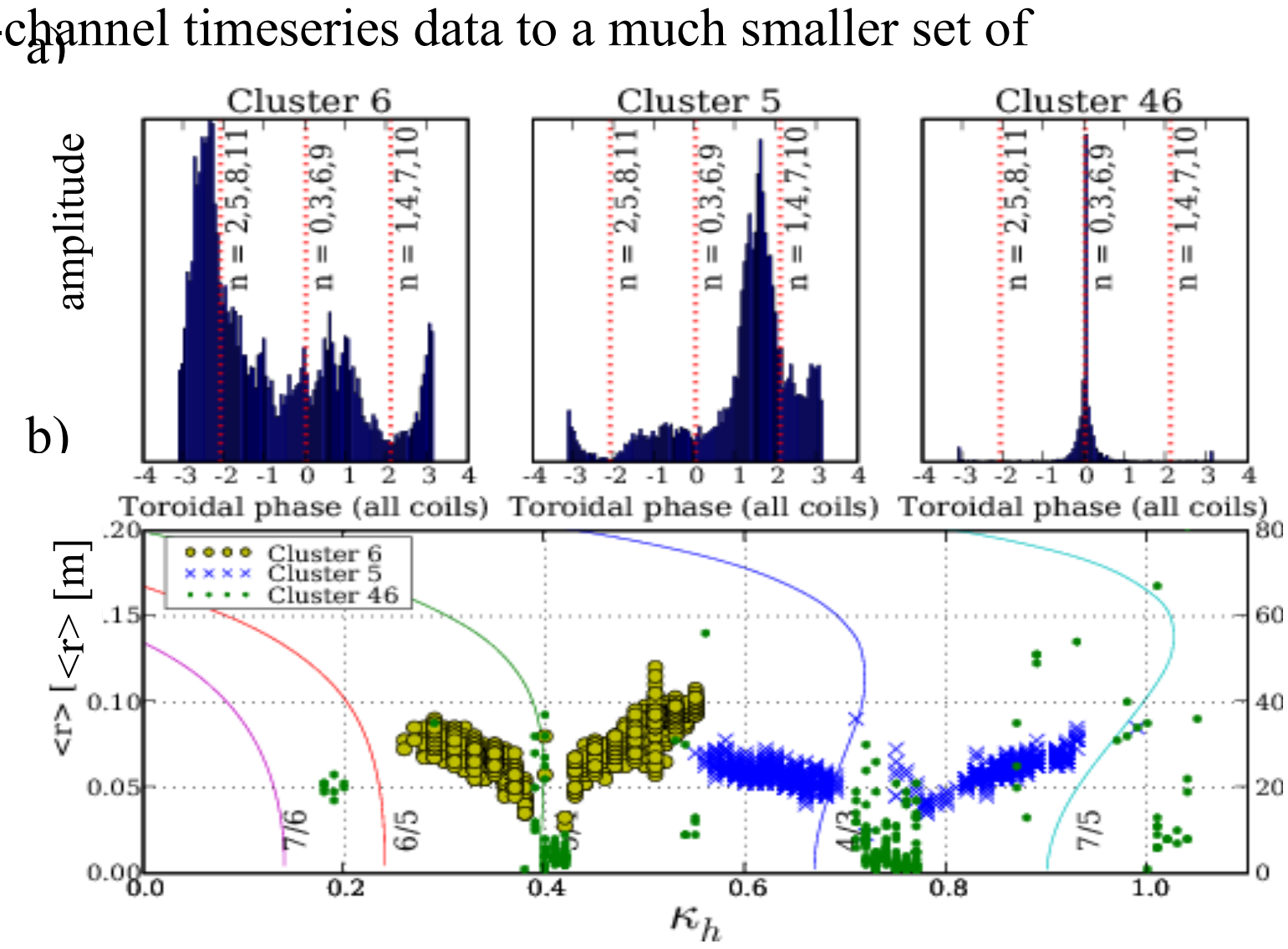

***Fig. 1**: Results of the classification of the configuration scan ($k_h$). Three of the clusters found are shown, Cluster 6 with mode numbers n/m=5/4, cluster 5 (n/m=4/3) and cluster 46 with n=0. In b), the mode frequency is shown, and thin lines are contours of rational rotational transform as a function of radius.*

To some degree this enables the analysis to adapt to magnetic coordinate systems which vary with the plasma configuration. This analysis is illustrated in Fig. 3, which shows three clusters of data points with distinct mode structure.

A significant class of fluctuations (such as clusters 5 and 6, Fig. 3) exhibit scaling which is Alfvénic with electron density (within a constant factor $\lambda$) and shear Alfvénic in rotational transform.

### 2.1 Interpretation

In low shear configurations near (but not at) resonance, global Alfvén eigenmodes (GAEs) [7] are predicted to cluster in the spectral gap $0<\omega<|k_{\|}V_A|$ which decreases as the transform approaches resonance ($\iota$ = 5/4 and 4/3), and which would lead to minima in f as follows. The Alfvén resonant frequency for the low positive shear typical of H-1 is approximately constant near the axis, and rises steeply toward the plasma edge. Using periodic boundary conditions to close the torus, then $k_{\|} = (m/R_0)(\iota - n/m)$, so $\omega \rightarrow 0$ linearly in the vicinity of a resonance ($\iota$ = n/m). Observed frequencies are proportional to $\omega/V_A = k_{\|} = (m/R_0)(\iota - n/m)$ and, by virtue of the linear dependence on ($\iota$ - n/m) either side of the minimum (~0) in $\omega$ at resonance (rational $\iota$ = n/m), show clear "V" structures near those rational surfaces. In addition to their intrinsic interest, in a low shear device such as H-1, it will be shown below that these can provide an accurate location of resonant surfaces under plasma conditions, which agree very well with recent magnetic field line mapping at high magnetic field.

Fig. 5 shows a more complete data set scaled to remove electron density variations, assuming Alfvénic scaling, and with lines for selected modes (n/m) showing the expected Alfvén frequency for a cylindrical model at the location of zero shear in the iota profile, which is approximately the condition for an Alfvén eigenmode. A better match to experimental data is found if this condition is modified so that when the corresponding resonance condition $\iota$ = n/m is not met at any radius (such as the right branch of the 5/4 mode : $0.4<k_h$), the iota value is taken at a radius <a>=15cm, near the plasma edge. Fig. 9 supports this; maximum fluctuation amplitude is further from the axis for the 5/4 range than the 4/3 range of modes. The identification of the 7/6 and 6/5 modes is not as certain as they are closer to the Nyquist spatial frequency of the 20 coil arrays when the "distortion" of the magnetic coordinate system is taken into account.

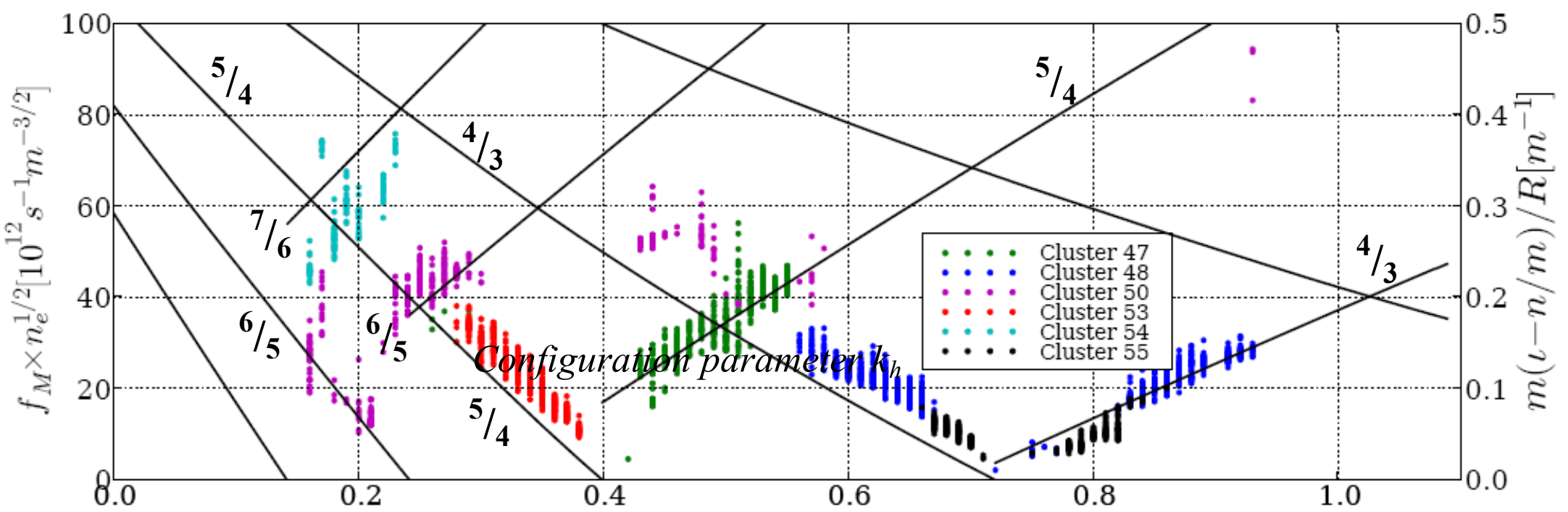

***Fig. 2:*** *A more complete data set re-scaled by $\sqrt{n_e}$ to show Alfvenic dependence on configuration parameter $k_h$. Lines show expected Alfven frequency at the stationary point in rotational transform profile when the correspoding resonance is in the plasma, and at a fixed radius (<a>~15cm) if not.*

The poloidal phase variation shown in Fig. 7 confirms the mode number m=4, and the approximately linear dependence of phase indicates a rotating mode in the ion diamagnetic direction. Although many of the identified modes have this rotation direction, the linearity of the phase is not as consistent as in Heliotron J using the same analysis. This is possibly a consequence of the more marked departure from circularity of the cross-section of H-1.

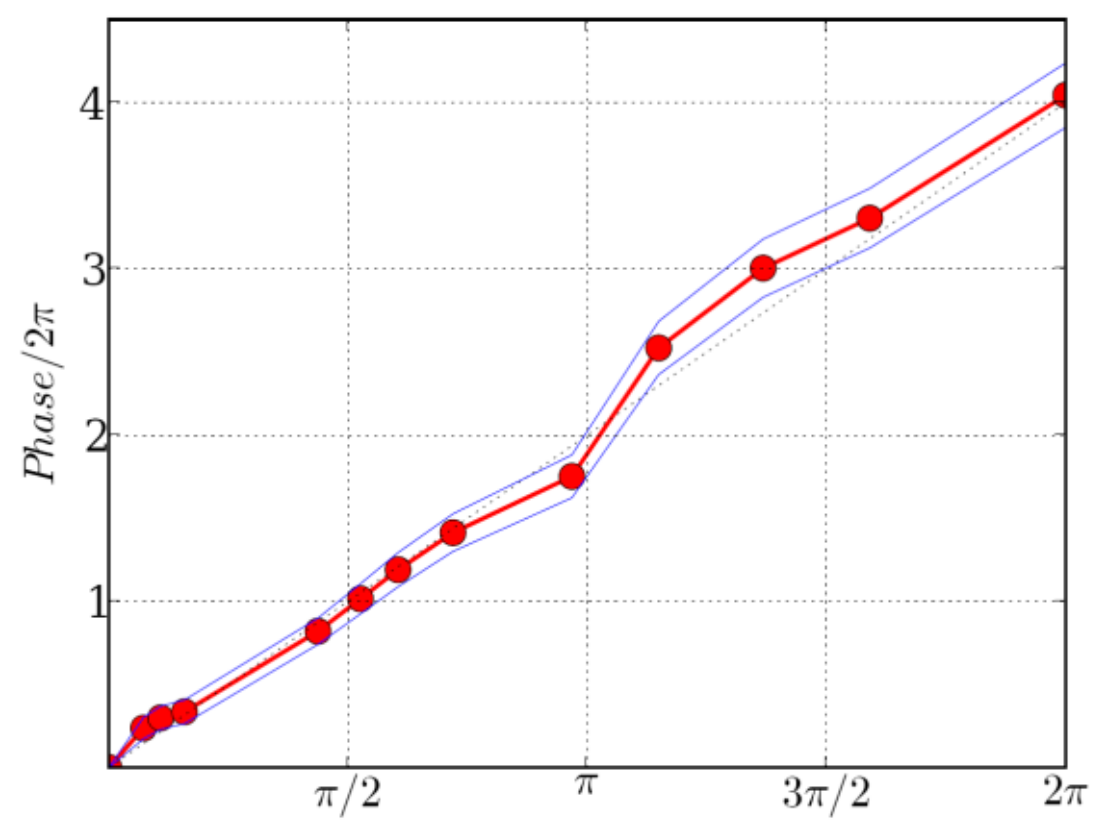

***Fig. 3:*** *Phase variation of the modes in cluster 6 with poloidal angle, showing a mode number m=4. The red line and points show the cluster means, and the thin blue lines the standard deviation within the cluster.*

Although the dependence on iota (ɨ - n/m) is clear, absolute frequency values are lower than predicted by a constant scale factor λ of 1/3. This could be caused by the effective mass density being higher than that of the constituent gas mix due to impurities or momentum transfer to background neutrals, but the extent of both these effects is expected to be too small to explain the entire factor. There is a report[8] of an instability driven by fast particles travelling at velocity reduced by a comparable scale factor (1/3) and an observation[9] of a toroidal Alfvén eigenmode with spectral components at 1/3 the expected frequency, but the physical mechanism is not clear, and the experimental conditions are somewhat different.

Initial computational studies using CAS3D[10] indicate that the 3D global modes and associated continua (Fig. 6) are largely unchanged in frequency from the simple cylindrical model used in the above analysis, with the exception of modes near gaps (HAE in Fig. 6) and those near zero frequency. These modes are up-shifted by a β-induced gap ~5 kHz, in spite of the low β ~ 0.018%. This frequency is typical of the lower limit of modes observed with clearly resonant mode numbers. In Fig. 4 for example, although the observed frequencies approach zero near the 4/3 and 5/4 resonances, the modes lose their m=3 or 4 character below about 10 kHz and show very much reduced poloidal variation in phase. The nature of the mode is more electrostatic, and together with the mode number change, is suggestive of the Beta-induced Alfvén Eigenmode (BAE) or the related Geodesic Acoustic Mode (GAM).

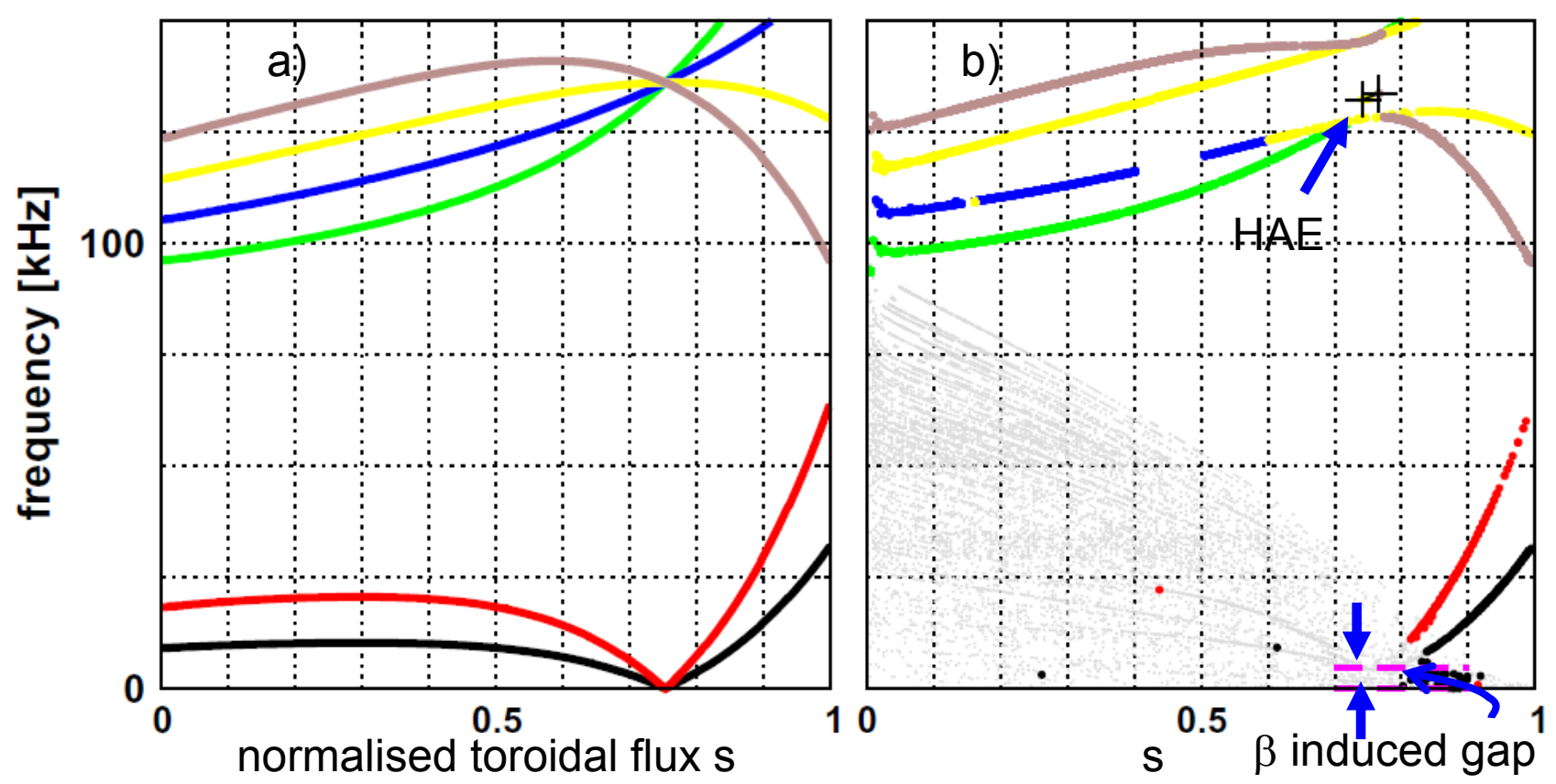

***Fig. 4:*** *CAS3D cylindrical (a) and 3D (b) frequency spectra*

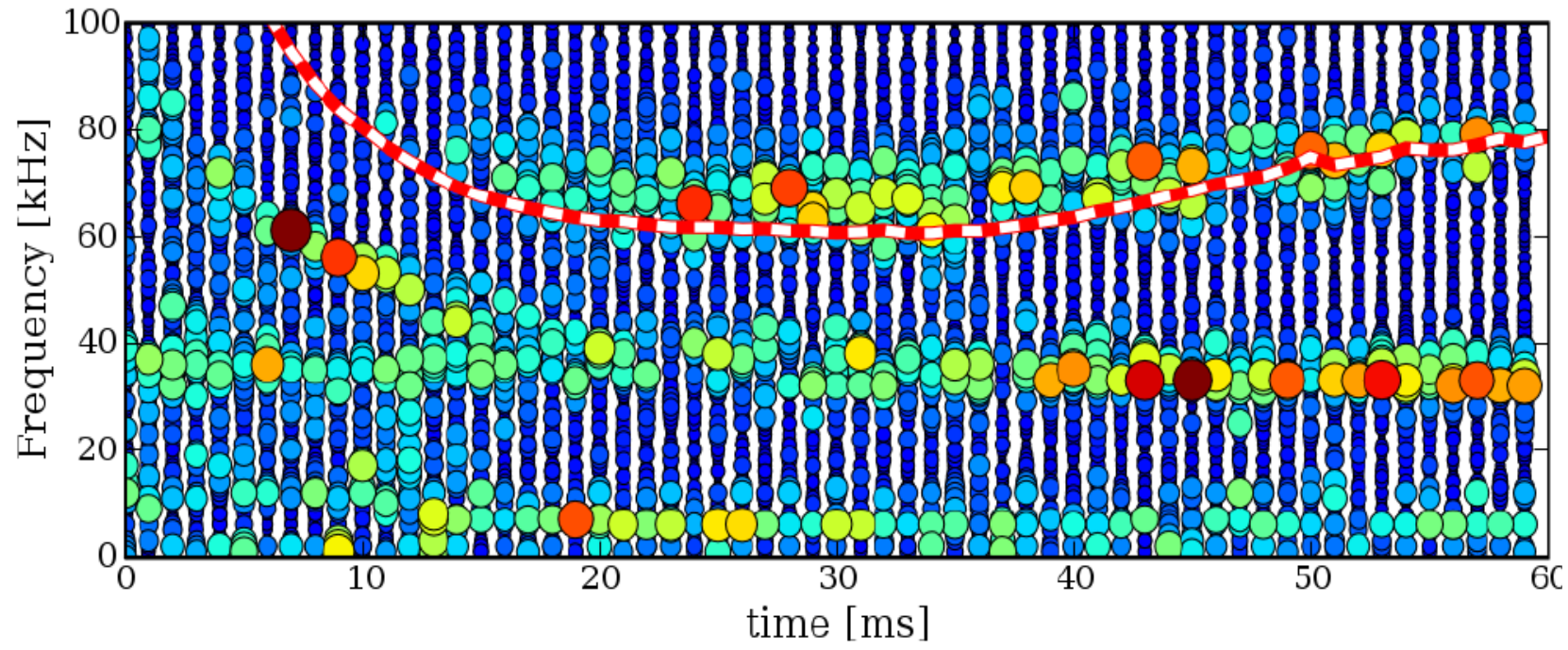

***Fig. 6:*** *Uppermost mode cluster (red dashes) exhibits Helical Alfvén Eigenmode scaling as density varies in time for one plasma discharge.*

At rotational transform approaching 1.4-1.5, there is a cluster of modes (Fig. 8) which are non-resonant, and with frequency dependence which matches that expected from another Helical Alfvén Eigenmode (HAE), but with a scale factor λ much closer to unity (0.85) . Although encouragingly close to unity, this is inconsistent with the resonant modes in Fig. 5.

## 2.2 Radial Mode Structure

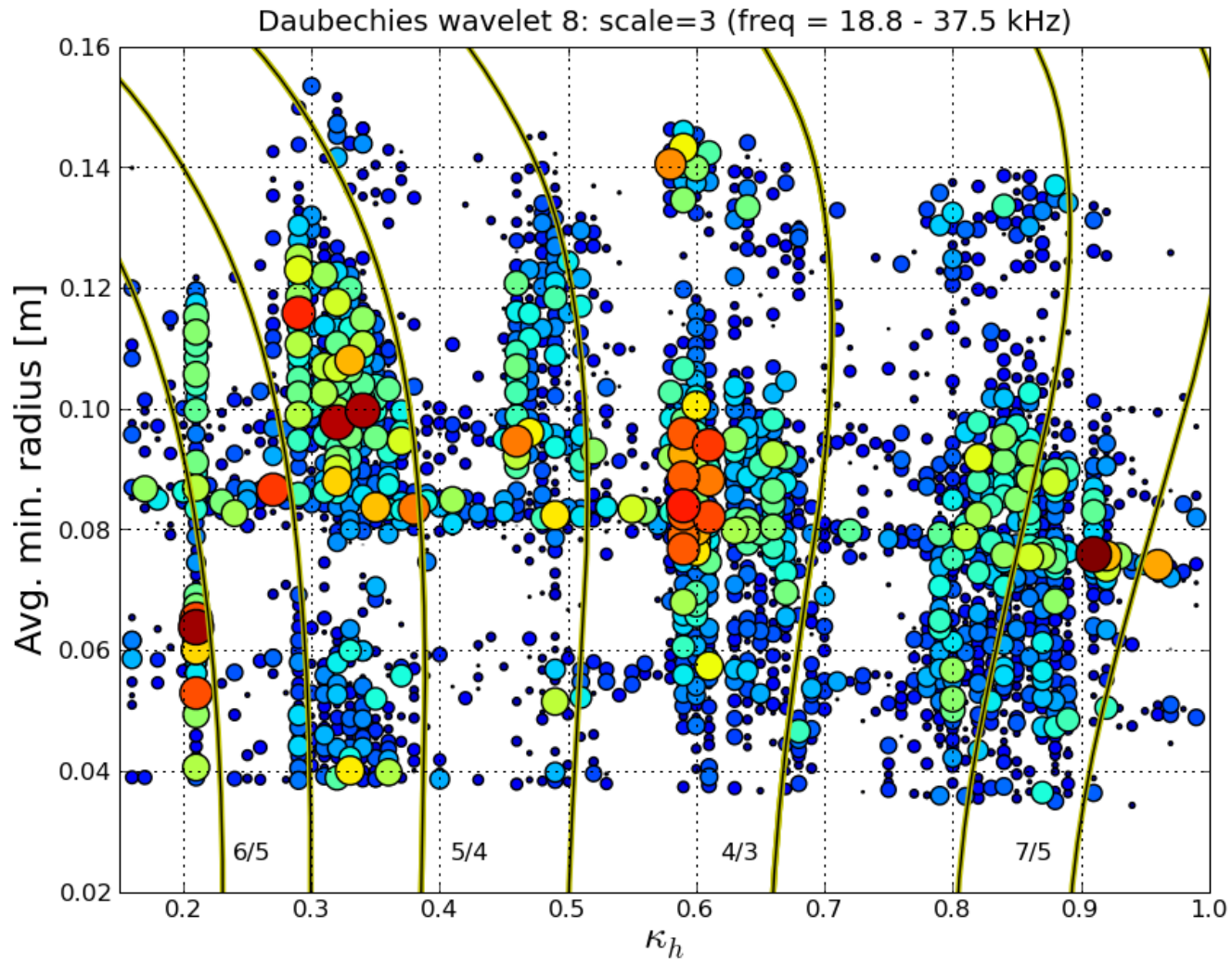

***Fig. 5:*** *Dependence of line-averaged density fluctuation amplitudes (dot size) on effective radius of the line of sight and configuration parameter $k_h$. The swept nature of the diagnostic causes reproducible global modes (e.g. $k_h$~0.6) to be artificially broken up into separate data points.*

The radial localisation of the fluctuations has been investigated with a precision 2mm interferometer and a 1.5mm fast scanning interferometer[11]. Fig. 9 shows the chordally integrated fluctuation amplitude as a function of configuration parameter $k_h$. The mode near the 5/4 resonance is seen to exist at $r < r(\iota=5/4)$ for $0.29 < k_h < 0.38$, while the mode near $k_h$~0.5 has a radial fluctuation profile consistent with localisation in the zero-shear region. Note that a mode localised at $r_x$ in the plasma should appear as $0<r<r_x$ in Fig 8. Mode structure simulations assuming m=3 and 4 are consistent with data and the explanation of the Alfvén eigenmodes given in §2.1. Fig. 10 shows the structure of the n=5, m=4 mode extracted by synchronous detection of the density fluctuations detected by a low-noise 2mm interferometer scanned through the plasma on a shot to shot basis. The figure shows the reproducibility of the results, and a qualitative correspondence with a simulation of an m=4 mode in the magnetic geometry of the heliac.

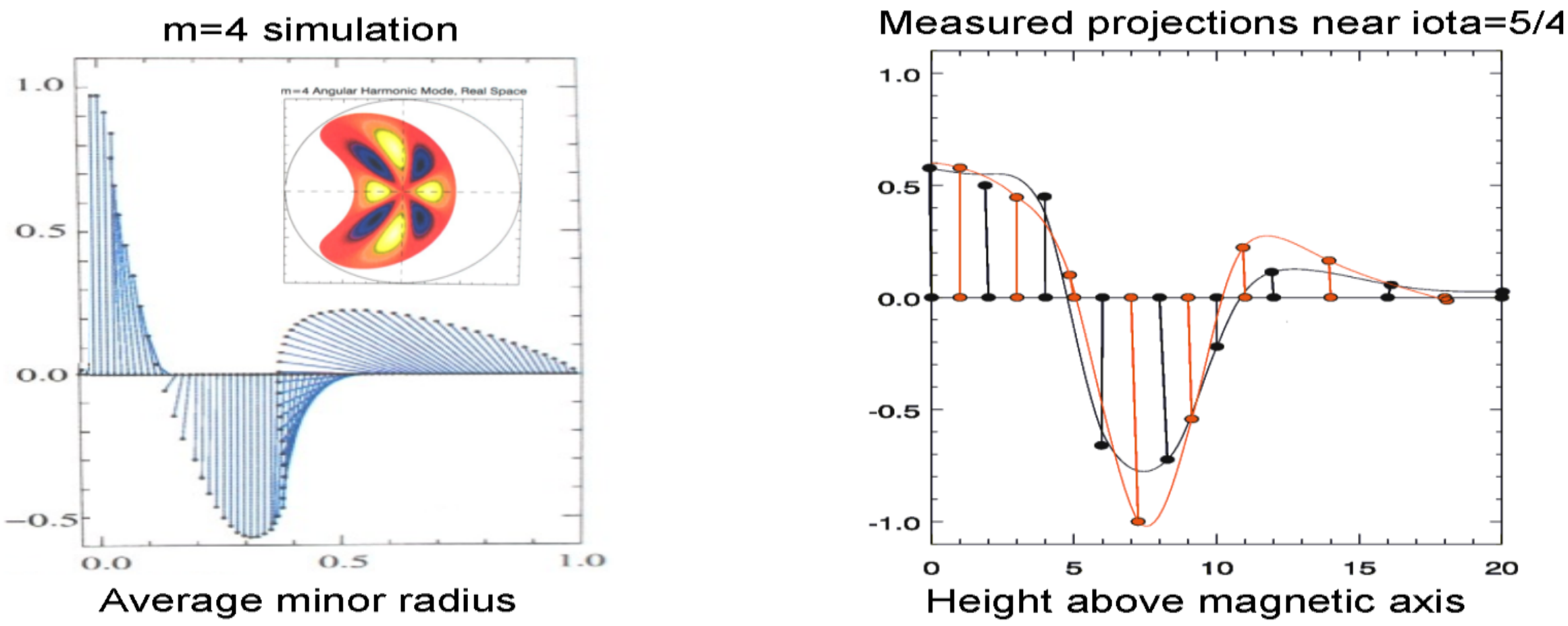

***Fig. 7:*** *Simulated and measured mode structure in line-integral density for the stronger modes near iota=5/4 (large circles in Fig. 8). The tilt of the phasors represents the phase angle. Chordal integration and symmetry causes phases to tend 0 or π even for rotating modes.*

The complexity of the plasma shape (**Error! Reference source not found.**, 2b), and its variation with configuration creates problems in analysis, possibly broadening the poloidal mode spectrum more than expected from toroidal and helical coupling. The variable plasma – probe distance makes mode localisation difficult. Consequently, the identification of ballooning modes by comparing mode amplitudes in regions of favourable and unfavourable curvature is difficult. However there are clearly some modes which are quite different to the Alfvén eigenmodes described above, either more broadband, or with a density dependence that is clearly non-Alfvénic. The latter distinction is more obvious when the plasma density varies during a single plasma pulse, and the frequency variation is clearly not $1/\sqrt{n_e}$.

The drive for these modes is not clear; fast particle driving sources are under investigation, and include both fast electrons and minority-heated H ions. At this time, no clear evidence of suitable fast particles has been found. It is unlikely that H or He ions in H-1 could reach the energy required to match the Alfvén velocity, and there is no obvious spectral indication of high energy components either through charge exchange to H atoms or acceleration of He ions by drag from fast H ions. However both these processes are indirect, and Doppler broadened features due to such high energies would be far into the wings of the spectral lines, and may be difficult to distinguish from the background. There is a distinct possibility of acceleration of either species by the high potential RF (~kV) on the antenna, and it is common to find an elevated temperature in either or both electron and ion species near the edge in RF heated H-1 plasma[12]. Recently observations of excitation of Alfvén eigenmodes by steep temperature gradients, in the absence of high-energy tails, have been reported.[13]

### 3. Effects of Configuration on Magnetic Surfaces

The plasma density falls dramatically near the resonances discussed in §2.1, leading to the question: To what extent is this due to the effect of magnetic islands? Magnetic islands are expected to be relatively less important in a heliac than in conventional stellarators because of the combination of flexibility and moderate shear – the shear is small enough so that in most configurations, there are no low order rationals present, but sufficiently large so that island width is reduced in configurations containing significant rationals. We have developed a uniquely accurate magnetic field mapping system[14] using tomography of electron beam currents collected by a rotating wire wheel, and a comprehensive model of the vacuum magnetic field in H-1. The model includes details of coils and busses, all significant sources of magnetic field error and stray fields. The computation and electron beam mapping measurement agree well and the number of transits is sufficient for measurements of iota to better than 3 decimal places in the vicinity of moderate order resonances (m,n ~ 8-15), using the computer fit only to second order.

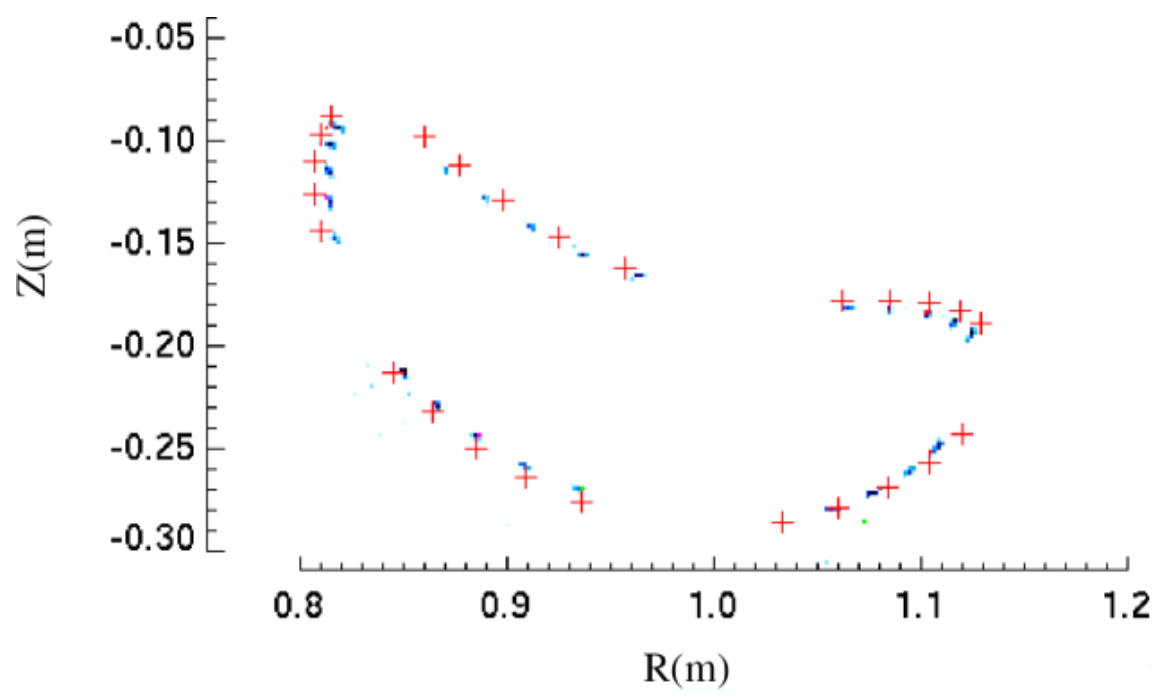

***Fig. 8: Measured vacuum punctures[ ] and computation[+].***

This allows comparison with the iota inferred from the Alfvén mode frequency measurements reversing the procedure outlined in §2. The agreement is good (δɨ/ɨ ~0.4%) when compared to direct electron beam measurements at low magnetic field.

Recently the mapping system was upgraded [15] to allow scans fast enough to allow partial imaging at full operating magnetic field. This showed a small change in iota due to helical distortion of conductors under the magnetic force loading, which when extrapolated to the conditions of Fig. 12b, reduces the discrepancy to δɨ/ɨ <0.2%. This demonstrates the potential of this technique in measurement of rotational transform in the presence of plasma, provided the device has low shear, or the radial location of the mode can be clearly identified.

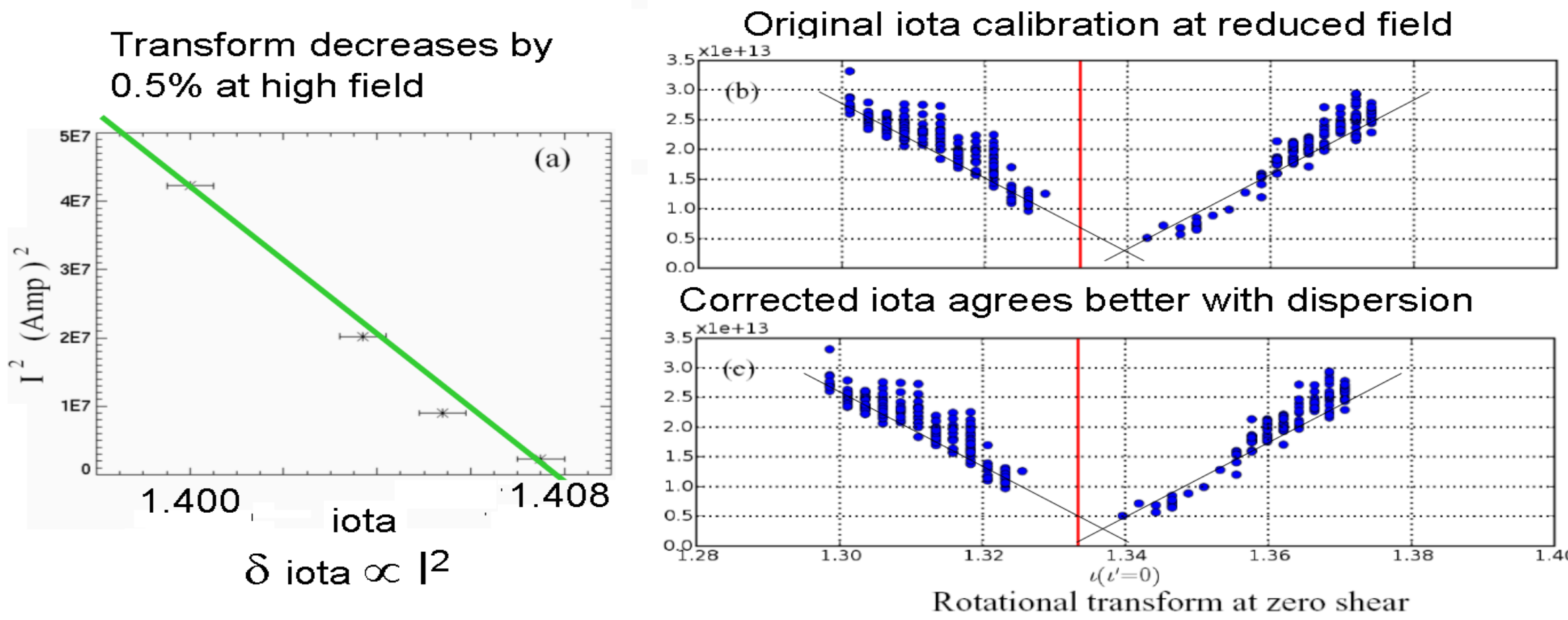

***Fig. 9 :*** *Comparison of transform determination using Alfvén eigenmode resonance and direct ebeam mapping(a). The discrepancy between the transform obtained from the symmetry point in the "V" structure of the observed frequency (b) and the computed transform value is halved (c) if the computed transform is corrected for a small distortion in the magnetic field coils due to the magnetic forces inferred from the results (a) of electron beam mapping at high field.*

Extensive tomographic configuration mapping and corresponding magnetic field computation shows that magnetic islands do not dominate the configuration except at iota ~ 1, and to a smaller degree at iota~3/2. The poor confinement near some resonances (**Error! Reference source not found.**a), especially near iota values of 5/4 and 4/3, seems therefore to be not simply due to destruction of magnetic surfaces by islands.

To allow a detailed investigation of magnetic islands, and their effect on plasma confinement, the iota~3/2 configuration was chosen because the islands inherent in the shape (elongation/indentation) of the heliac, are large enough that an effect could be expected, but not so large as to totally destroy confinement. Experiments in the vicinity of iota ~3/2 were performed in Argon plasma over a range of parameters, and do not show any clear degradation in confinement or any noticeable features at the island position (δ), as measured by Langmuir probe estimates of density and temperature. However, for lower neutral densities, there is a small increase in confinement within the island (Fig. 13), and a steepening of the potential profile in the vicinity of the core. Investigations into similarity with core electron root enhanced confinement[16] are ongoing.

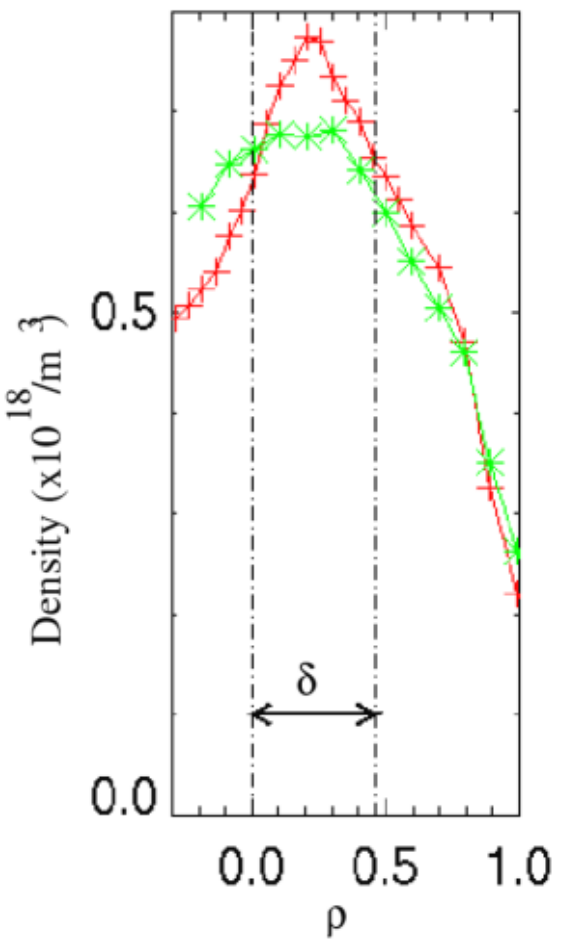

*Fig. 10:* *Plasma density increase in presence of an island (indicated by δ) for initial neutral densities of 0.8(+) and 1.6(✳) $10^{18}m^3$*

## Conclusion

We have presented strong evidence for Alfvénic scaling of magnetic fluctuations in H-1, in $n_e$, iota and ω. There is however unclear scaling in B, and an unexplained factor of ~3 in the frequency of near-resonant modes. In low shear configurations, we have shown that near-resonant Alfvén eigenmodes can provide a sensitive iota diagnostic. The increased dimensionality of parameter space arising from such systematic investigations is efficiently handled by datamining. This technique is being applied to a number of devices internationally, coordinated through an open source software project[4]. The understanding provided by this unique combination of flexibility and variety of advanced diagnostics can contribute to the understanding of Alfvénic activity, and the effects of magnetic islands in present-day and planned extremely energetic fusion plasmas.

## Acknowledgements

The authors would like to thank the H-1 team for continued support of experimental operations as well as M. Hole, B. McMillan and T. Luce for useful discussions. This work was performed on the H-1NF National Plasma Fusion Research Facility established by the Australian Government, and operated by the Australian National University, supported in part by Australian Research Council Grants DP0344361 and DP0451960 and by the U.S. Department of Energy under Contract No. DE-AC05-00OR22725 with UT-Battelle, LLC.

## References

[1] Hamberger S.M., Blackwell B.D., Sharp L.E. and Shenton D.B. “H-1 Design and Construction”, *Fusion Technol*. **17**, 1990, p 123.

[2] Harris, J.H., Cantrell, J.L. Hender, T.C., Carreras, B.A. et al. *Nucl. Fusion* **25**, 623 (1985).

[3] Harris, J.H., Shats, M.G., Blackwell, B.D., Solomon, W.M., et al. *Nucl. Fusion,* **44** (2004) 279-286.

[4] Pretty D.G, Blackwell B.D., Harris J.H. “A data mining approach to the analysis of Mirnov coil data from a flexible heliac”, submitted to Computer Physics Communications and open source code http://code.google.com/p/pyfusion/.

[5] Dudok de Wit T, Pecquet A L, Vallet JC, and Lima R. *Phys Plasmas*, **1**(10):3288-3300, 1994

[6] Witten Ian H, and Frank Eibe. Data Mining: *Practical machine learning tools and techniques.* Morgan Kaufmann, 2nd edition, 2005.

[7] Weller, A., Anton M., et al., *Physics of Plasmas*, 8 (2001) 931-956 and Wong K.L., “A review of Alfvén eigenmode observations in toroidal plasmas” *Plasma Phys. and Contr. Fusion* **41** (1): r1-56 1999

[8] Biglari H., Zonca F., and Chen L., “On resonant destabilization of toroidal Alfvén eigenmodes by circulating and trapped energetic ions/alpha particles in tokamaks” *Phys. Fluids* B 4 **8** 1992, and S Ali-Arshad and D J Campbell, “Observation of TAE activity in JET” *Plasma Phys. Control. Fusion* 37 (1995) 715-722.

[9] Maraschek, M., Günter S., Kass T., Scott B., Zohm H., and ASDEX Upgrade Team "Observation of Toroidicity-Induced Alfvén Eigenmodes in Ohmically Heated Plasmas by Drift Wave Excitation", *Phys. Rev. Lett.* **79** (1997) 4186-9

[10] C. Nuhrenberg. "Computational ideal MHD: Alfvén, sound and fast global modes in W7-AS". Plasma Phys. Control. Fusion, 41(9):1055–70, 1999.

[11]Howard J. and Oliver D., "Electronically swept millimetre-wave interferometer for spatially resolved measurement of plasma electron density", *Applied Optics* **45** (2006) pp. 8613-8620

[12]Michael C. A., Howard J. and Blackwell B. D., "Measurements and modeling of ion and neutral distribution functions in a partially ionized magnetically confined argon plasma", *Physics of Plasmas* 75, (2004) 4180-4182.

[13]Nazikian R. et al. Phys. Rev. Lett. 96, (2006) 105006.

[14] Kumar, S.T.A, Blackwell, B.D. and Harris, J.H., "Wire tomography in the H-1NF heliac for investigation of fine structure of magnetic islands", Rev. Sci. Ins, 78 (2007) 013501-8

[15]Kumar, S.T.A and Blackwell, B.D, "Accurate determination of the magnetic geometry of the H-1NF heliac", submitted to *Nuclear Fusion*.

[16] Yokoyama M., Maasberg H., Beidler C.D.,.et al. "Core electron-root confinement (CERC) in helical plasmas." Nucl. Fusion, 47(2007):1213–1219.